\documentclass[
aps, 
prl,
preprintnumbers,
twocolumn,
superscriptaddress,
nofootinbib,
floatfix,
10pt
]{revtex4-2}

\usepackage{graphicx}
\usepackage{dcolumn}
\usepackage{bm}
\usepackage{float}
\usepackage{braket}
\usepackage{amsmath}
\usepackage{amssymb}
\usepackage{mathtools}
\usepackage{comment}
\usepackage{xcolor}
\usepackage[bottom]{footmisc}
\usepackage{hyperref}

\def\orcid#1{\kern .08em\href{https://orcid.org/#1}{\includegraphics[keepaspectratio,width=0.7em]{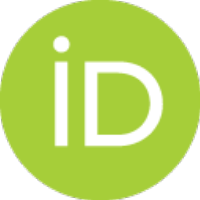}}}

\newcommand{\defeq}{\vcentcolon=}

\begin{document}

\title{\textbf{An Eigenvector-Continued Green's Function Method} 
}

\author{Stefano~Brolli\orcid{0009-0005-8290-8462}}
 \email{Stefano.Brolli@unimi.it}
\affiliation{Dipartimento di Fisica “Aldo Pontremoli,” Universit\`a degli Studi di Milano, via Celoria 16, I-20133 Milano, Italy}
\affiliation{INFN, Sezione di Milano, via Celoria 16, I-20133 Milano, Italy}

\author{Carlo~Barbieri\orcid{0000-0001-8658-6927}}
 \email{Carlo.Barbieri@unimi.it}
\affiliation{Dipartimento di Fisica “Aldo Pontremoli,” Universit\`a degli Studi di Milano, via Celoria 16, I-20133 Milano, Italy}
\affiliation{INFN, Sezione di Milano, via Celoria 16, I-20133 Milano, Italy}

\date{\today}

\begin{abstract}
\noindent Emulating the Green's function using reduced-order equations is challenging because of the amount of information it contains. This information includes the one-particle spectroscopy over the entire energy spectrum, the one-body density matrix, and the ground-state energy. Nevertheless, the computational speedup offered by emulators is essential for applications requiring repeated model evaluations, which are ubiquitous in many-body physics. These include parameter calibration, uncertainty quantification, phase-diagram mapping, and geometry optimization. In this Letter, we introduce a systematically improvable emulator based on eigenvector continuation and moment-matching. The resulting approximation, which we refer to as an ``eigenvector-continued Green's function,'' can be obtained by solving linear equations, enabling an efficient separation of the computational workload into offline and online stages.
\end{abstract}

\maketitle

The single-particle Green's function provides a wealth of information on the structure and dynamics of interacting systems. Its spectral representation encodes the full one-particle spectroscopy, enabling studies of nuclear spectra~\cite{brugnara2025, Cipollone2015}, electronic excitations in condensed matter systems~\cite{Hybertsen1986}, and molecular ionization energies~\cite{Cederbaum1978}, while also providing access to the one-body density matrix and ground-state energy~\cite{Dickhoff2004}. The Green's function also encodes scattering wavefunctions~\cite{Capuzzi1996, cederbaum2001}, offering a description of electron- and positron-molecule scattering~\cite{Meyer1989, Rawlins2023} and a promising route toward \emph{ab initio} nuclear optical potentials~\cite{Brolli2026, Brolli2025, Idini2019}. However, state-of-the-art Green's function calculations are computationally expensive~\cite{Brolli2026, Brolli2025, Soma2020, Govoni2015, VanHoucke2012}, placing any application that requires multiple evaluations out of reach. This issue can be resolved by reduced-order models that require only a fraction of the computational cost, however, the amount of information contained in the Green’s function makes this route challenging. In particular, wavefunction-based methods, such as eigenvector continuation (EC)~\cite{Frame2018}, target the ground-state of the system and do not directly contain information on scattering and spectroscopy. This shortcoming is particularly evident as the need for repeated evaluations arises across many areas of quantum many-body physics, from Bayesian uncertainty quantification~\cite{Konig2020, Hu2022, Svensson2024} and the calibration of low-energy constants (LECs)~\cite{Hu2025:N3LOTexas} in nuclear physics to geometry optimization and molecular dynamics simulations in quantum chemistry~\cite{Rath2025}. More recently, Green's function emulation has also been explored in solid-state physics to reduce the computational cost of exploring parameter spaces for materials design and identifying phase boundaries~\cite{hogan2026}. In this Letter, we introduce an approach for emulating the zero-temperature Green's function in finite systems by enforcing exact moment-matching up to the EC approximation of the ground-state. To demonstrate the versatility of our approach, we benchmark its performance on the Richardson model of nuclear pairing~\cite{Richardson1963}, which is a widely adopted testbed for many-body methods in nuclear physics~\cite{Brolli2025, Hu2026, Franzke2024, Rigo2023}, and the Hubbard ring, which is more common in condensed matter physics~\cite{Lieb1968, Shastry1990, Boguslawski2014, Yao2010, Bannister2000} and has also been studied on quantum computers~\cite{Suchsland2022}. Both models exhibit an affine dependence on the interaction strength, a property inherited by the emulator that enables a decomposition of the computational effort into offline and online stages~\cite{Drischler2023}. Our emulator reproduces the ground-state energies obtained with EC and accurately reproduces states with large spectroscopic factors, which are typically probed in one-particle knockout and transfer reactions. In particular, the spectroscopic factor gives a lower bound on the fidelity with which that eigenstate can be represented by the emulator~\cite{SupplMat}. 

\begin{figure}
    \centering
    \includegraphics[width=\linewidth]{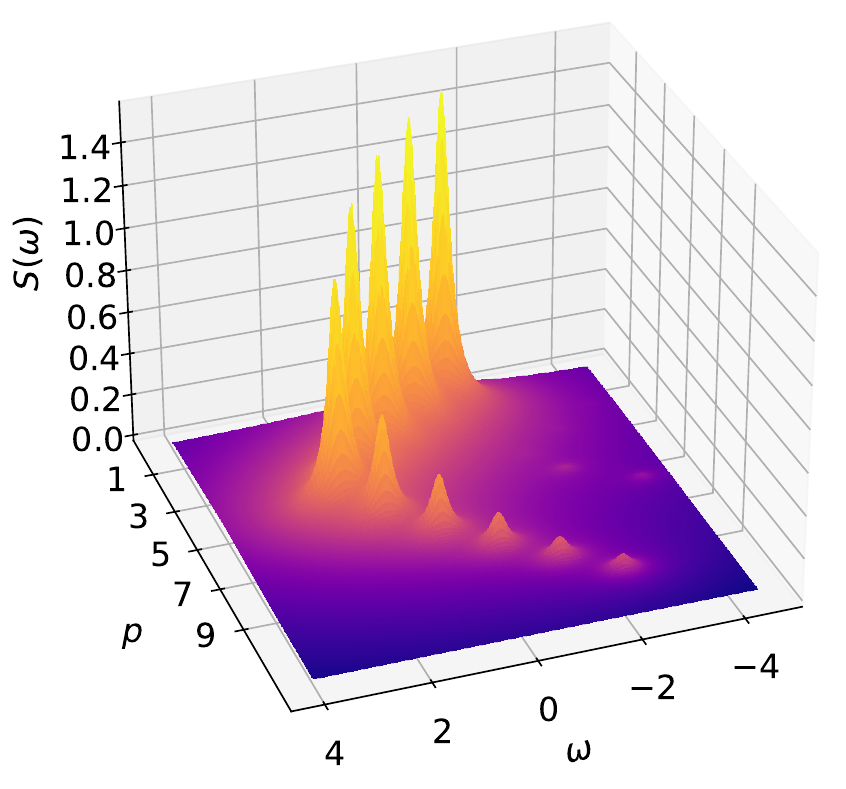}
    \caption{Emulated hole spectral-strength distribution for each single-particle level $p$ of the Richardson model at coupling $g=1.3$. The peaks were smeared with a Lorentzian of width $\eta=0.2$. The emulation was performed as detailed in the text.}
    \label{fig:SpecFunc_g1.3_3D_emul}
\end{figure}

{\itshape Emulator equations}---We consider a Hamiltonian with an affine dependence on a single parameter $\theta$,
\begin{equation} \label{eqHamiltonianGeneral}
    H = H_0 + \theta H_1.
\end{equation}
The following discussion can be naturally extended to the more general case of a Hamiltonian with an affine dependence on many parameters.
Our approach for emulating the Green's function starts by building the ground-state using EC. That is, we assume that the ground-state of the Hamiltonian~\eqref{eqHamiltonianGeneral} can be computed for $n_b \ll \dim(H)$ distinct values of the parameter $\theta$, denoted by $\{\theta_i\}_{i=1}^{n_b}$. These eigenvectors—often referred to as ``snapshots'' in the literature—span a subspace $\widetilde{\mathcal{H}}^A$ of the full $A$-particle Hilbert space that contains the dominant components of the ground-state wave function as $\theta$ varies. Throughout this work, we use a tilde to denote objects associated with the reduced-order models. Projecting the full Hamiltonian~\eqref{eqHamiltonianGeneral} onto the subspace spanned by the orthonormalized snapshots $\{ \ket{\psi_j} \}_{j=1}^{n_b}$ yields the reduced eigenvalue problem
\begin{equation}
    \left( \widetilde{H}_0 + \theta \widetilde{H}_1  \right) \ket{\widetilde{\Psi}_0^A(\theta)} = \widetilde{E}_0^A(\theta) \ket{\widetilde{\Psi}_0^A(\theta)},
\end{equation}
where $\widetilde{E}_0^A$ and $\ket{\widetilde{\Psi}_0^A}$ are the reduced-order approximations of the ground-state energy $E_0^A$ and wavefunction $\ket{\Psi_0^A}$, respectively. We also introduce the notation
\begin{equation}
    \ket{\widetilde{\Psi}_0^A} = \sum_j \beta_j \ket{\psi_j}
\end{equation}
to represent the decomposition of the reduced-order ground-state wavefunction on the orthonormalized snapshots $\{ \ket{\psi_j} \}_{j=1}^{n_b}$. 
To build an accurate approximation of the Green's function, we employ its Källén-Lehmann representation~\cite{Kallen1952, Lehmann1954}
\begin{equation} \label{eqLehmann}
    G_{\alpha \beta} (\omega) = \sum_k \frac{\mathcal{Y}_\alpha^k \left(\mathcal{Y}_\beta^k\right)^*}{\omega -\epsilon_k^- - i\eta} + \sum_n \frac{\left(\mathcal{X}_\alpha^n\right)^*  \mathcal{X}_\beta^n}{\omega -\epsilon_n^+ + i\eta}.
\end{equation}
Here, $\alpha$ labels a single-particle state of the computational basis, while $E_i^{A\pm1}$ and $\ket{\Psi_i^{A\pm1}}$ denote the energy and eigenvector, respectively, of the $i$-th eigenstate of the $(A\pm1)$-particle system. The quantities $\epsilon_k^- \defeq E_0^A - E_k^{A-1}$ and $\epsilon_n^+ \defeq E_n^{A+1} - E_0^A$ are the one-particle removal and addition energies associated with the $k$-th and $n$-th eigenstates of the $(A-1)$- and $(A+1)$-particle systems, respectively. The corresponding spectroscopic amplitudes are defined as $\mathcal{Y}^k_\alpha \defeq \braket{\Psi_k^{A-1}|c_\alpha|\Psi_0^A}$ and $\mathcal{X}^n_\alpha \defeq \braket{\Psi_n^{A+1}|c^\dagger_\alpha|\Psi_0^A}$. The Green's function provides access to the complete one-particle spectroscopy through the spectral-strength distribution
\begin{equation}
    S(\omega) \defeq \sum_n SF_n^+ \delta(\omega - \epsilon_n^+) + \sum_k SF_k^- \delta(\omega - \epsilon_k^-),
\end{equation}
where
\begin{equation}
\begin{split}
    SF_n^+ &{}\defeq \sum_\alpha \left\lvert \mathcal{X}^n_\alpha \right\rvert^2, \\
    SF_k^- &{}\defeq \sum_\alpha \left\lvert \mathcal{Y}^k_\alpha \right\rvert^2
\end{split}
\end{equation}
are the spectroscopic factors associated with particle addition and removal, respectively. This distribution encodes the excitation spectrum of the $(A\pm1)$-particle systems accessible through one-particle addition and removal, together with the corresponding spectroscopic strengths, which quantify the single-particle character of each state. 
We construct the Green's function by introducing reduced $(A+1)$- and $(A-1)$-particle subspaces, $\widetilde{\mathcal{H}}^{A \pm 1}_\nu$, generated starting from the states $\{ c_\alpha^\dagger \ket{\psi_j} \}_{j \alpha}$ and $\{ c_\alpha \ket{\psi_j} \}_{j \alpha}$, respectively. These subspaces are systematically enlarged while preserving the linear dependence of the projected Hamiltonian on $\theta$. To this end, we discuss two possible enrichment strategies. 

The first strategy, which we refer to as the ``power grouping technique,'' aims to preserve the moments of the spectral function for all values of $\theta$ within the accuracy of the EC approximation to the ground-state wavefunction. The $\mu$-th moment of the particle-addition or particle-removal spectral function is defined as
\begin{equation}
    m^{(\mu)}_{\pm} \defeq \sum_i (\epsilon_i^{\pm})^\mu SF_i^{\pm}.
\end{equation}
The information contained in the moments characterizes the distribution and fragmentation of the single-particle strength. The exact matching of moments up to order $2\nu+1$ relies on the power series expansion of $H = H_0 + \theta H_1$. By collecting terms of $H^\mu$ according to their powers of $\theta$, we obtain
\begin{equation}
    H^\mu = \sum_{k=0}^\mu \theta^k O_k^{(\mu)},
\end{equation}
where $O_k^{(\mu)}$, with $0 \leq k \leq \mu \leq \nu$, denotes the sum of all ordered products containing $k$ factors of $H_1$ and $\mu-k$ factors of $H_0$. By applying all operators $O_k^{(\mu)}$ up to $\mu = \nu$ to the initial particle-addition and particle-removal states, we systematically enrich the subspaces to capture moments up to $2\nu + 1$. The price to pay for the exact moment-matching (up to the EC wavefunction) for all $\theta$ is a quadratic growth in the dimension of the subspaces. 

In this Letter, we will focus on the second approach, namely a ``diagonal approximation technique," which provides a more gentle linear scaling by sacrificing exact moment-matching for all couplings and relying instead on moment-matching only at the snapshots. This is done by considering basis states of the form $\{ H^\mu(\theta_j) c^\dagger_\alpha \ket{\Psi_0^A(\theta_j)} \}_{j=1 \ldots n_b, \alpha, \mu = 0 \ldots \nu}$ and $\{ H^\mu(\theta_j) c_\alpha \ket{\Psi_0^A(\theta_j)} \}_{j=1 \ldots n_b, \alpha, \mu = 0 \ldots \nu}$ for the $A+1$ and $A-1$ sectors, respectively.  

After orthonormalization, we denote the resulting basis vectors by $\{ \ket{b_j^{A \pm 1}} \}$ and express the excited states in the reduced subspaces as
\begin{equation}
    \ket{\widetilde{\Psi}^{A\pm1}_i} \defeq \sum_{j} A^i_{j} \ket{b_j^{A \pm 1}}.
\end{equation}
 
We now specialize our discussion to the particle sector; however, the hole sector can be treated analogously~\cite{SupplMat}. The coefficients $A^n_{j \alpha}$ can be calculated by solving the eigenvalue problem (EP)

\begin{equation}  \label{eqECParticleSpace}
    \widetilde{H}^{A+1} \bm{A}^n = \widetilde{E}_n^{A+1} \bm{A}^n,
\end{equation}
where 
\begin{equation}
    \begin{split}
        \left[ \widetilde{H}^{A+1} \right]_{k j} &{}\coloneqq  \braket{b_k^{A+1}| H | b_j^{A+1}}, \\
        \left[ \bm{A}^{n, +} \right]_{j \alpha} &{}\coloneqq A_{j\alpha}^{n, +}.
    \end{split}
\end{equation}
The spectral amplitude of each pole is calculated from its definition according to
\begin{equation}
    \widetilde{\mathcal{X}}^n_{\alpha} = \sum_{k \delta} \sum_j  \left(A^{n, +}_{k} \right)^*  \widetilde{N}^{A+1}_{k, j\alpha} \beta_{j},
\end{equation}
where
\begin{equation}
        \left[ \widetilde{N}^{A+1}  \right]_{k, j \alpha} \coloneqq \braket{b_k| c^\dagger_\alpha | \psi_j}.
\end{equation}
The Green's function can then be reconstructed from the Källén-Lehmann representation \eqref{eqLehmann}. We refer to the Green's function built from this approach as an eigenvector-continued Green's function (ECGF).

The EP in Eq.~\eqref{eqECParticleSpace} has a maximum dimension of $\tfrac{1}{2} n_b D (\nu + 1) (\nu + 2)$ when employing the power grouping technique and $n_b D (\nu + 1)$ when using the diagonal method instead. Here, $D$ denotes the number of single-particle states in the computational basis. The dimension of the EP can be reduced by removing linearly dependent components and by exploiting the symmetries of the Hamiltonian, which are readily inherited by equation~\eqref{eqECParticleSpace}.

{\itshape Accuracy of the emulator}---To quantify the accuracy of the emulator, we employ the Wasserstein-1 ($W_1$) distance~\cite{Vallender1974, Clark1984, peyre2020}, defined as
\begin{equation}
    W_1 \defeq \int_{-\infty}^{+\infty} d\omega \ \left\lvert \overline{CS}_{\mathrm{emul}}(\omega) - \overline{CS}_{\mathrm{exact}}(\omega) \right\rvert.
\end{equation}
Here, $\overline{CS}(\omega)$ denotes the normalized cumulative spectral distribution~\cite{SupplMat}. The $W_1$ distance is also known as the earth mover's distance, as it measures the minimum cost of transporting spectral-strength along the energy axis to transform the emulated distribution into the exact one, with the cost being proportional to the amount of strength displaced and the distance over which it is transported.

Throughout this Letter, we show the spectral functions with a finite broadening $\eta = 0.1$–$0.2$ to facilitate visualization. However, the physical limit for the systems considered here corresponds to $\eta \to 0^+$, where the spectral functions become sums of Dirac delta distributions. In this limit, $L^p$ norms are poorly suited to quantifying the agreement between spectral functions. In fact, even arbitrarily close poles have no overlap unless their energies coincide exactly, making such norms insensitive to the proximity of the corresponding spectral features. In contrast, the $W_1$ distance remains well defined in the limit $\eta \to 0^+$ and provides a meaningful measure of the differences in both the positions and strengths of the spectral peaks.

{\itshape Richardson model}---The Hamiltonian of the Richardson model is \cite{Richardson1963}
\begin{equation}
    H = \sum_{p=1}^D \sum_{\sigma = \uparrow, \downarrow} (p-1) c^\dagger_{p \sigma} c_{p \sigma} - \frac{g}{2} \sum_{p, q=1}^D c_{p \uparrow}^\dagger c_{p \downarrow}^\dagger c_{q \downarrow} c_{q \uparrow}.
\end{equation}
It represents a system of fermions with $D$ equally spaced energy levels and a constant pure-pairing interaction. We will always consider the case of $D=10$ energy levels at half-filling. The Green's function of the Richardson model is diagonal both in the single-particle level and spin, and it is invariant under spin-flip~\cite{Brolli2025}. This allows a factorization of equation
\eqref{eqECParticleSpace} into $D$ EPs of size $n_b$.  We consider snapshots taken at the following values of the coupling $g = \{ -0.3, -0.1, 0.1, 0.3, 0.5 \}$ and truncate the expansion of the ($A\pm1$)-particle subspaces at order zero, imposing $\nu = 0$. 
With these considerations, evaluating the ECGF requires the diagonalization of ten $5 \times 5$ matrices in the particle sector, ten $5 \times 5$ matrices in the hole sector, and one $5 \times 5$ matrix for the EC ground-state. By comparison, the full-order calculation requires the diagonalization of twenty $126 \times 126$ matrices (ten for each of the particle and hole sectors) and one $256 \times 256$ matrix for the ground-state. Due to the symmetry between particles and holes, we focus on the hole-part of the Green's function. 
\begin{figure}
    \centering
    \includegraphics[width=\linewidth]{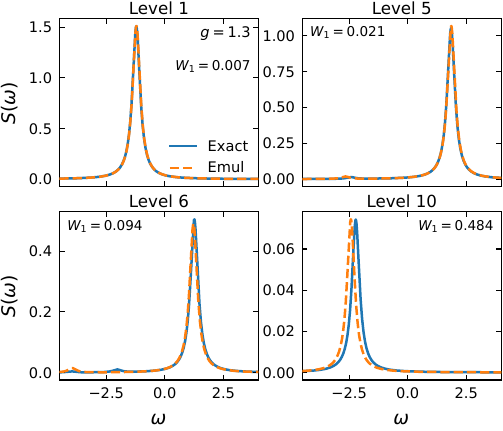}
    \caption{Dominant quasihole peaks of the spectral-strength distribution of the Richardson model for the coupling $g=1.3$. Note the different scales for the spectral strength on the vertical axes. The panels show the levels $p=1, 5, 6,$ and $10$. The plot shows the exact and emulated peaks, with a Lorentzian smearing of width $\eta=0.2$ of the Dirac delta distributions. Each panel also reports the $W_1$ distance of the exact and emulated distributions.}
    \label{fig:SpecFunc_g1.3_partial_waves}
\end{figure}
Figure \ref{fig:SpecFunc_g1.3_3D_emul} shows the emulated hole spectral-strength distribution of the Richardson pairing model at $g=1.3$, corresponding to a substantial extrapolation into the strong-coupling regime. We adopt a smearing width of $\eta=0.2$ for visualization purposes. Notably, the Richardson model undergoes a transition to the superfluid regime at approximately $g \approx 0.68$~\cite{Franzke2024}. Thus, the results presented here demonstrate the ability of the emulator to extrapolate well into the superfluid regime. Figure \ref{fig:SpecFunc_g1.3_partial_waves} presents a benchmark against the exact hole spectral distribution strength in the region around the largest peaks of the Richardson model. The figure shows the closest and farthest levels from the unperturbed Fermi level $F=5.5$. The large peaks appearing at levels below $F$ are reproduced with near-perfect agreement, despite the small dimension of the reduced EP given by the choice $\nu = 0$ and the large gap from the closest training point. The $W_1$ distance of these components is $\sim 10^{-2}$, proving the accuracy of the description. This is a consequence of having a dominant quasihole peak with a large spectroscopic factor ($95\%$ and $66\%$ for $p=1$ and $p=5$, respectively): these are mathematically guaranteed to be described well by the emulator~\cite{SupplMat}. The largest quasihole peaks associated to levels above the Fermi surface are instead complex satellite fragments found at energies below the Fermi energy. These carry a much smaller strength of $32\%$ and $4.7\%$ for the $p=6$ and $p=10$ levels, respectively. Nevertheless, these peaks are predicted by the emulator, even if their energy is not as accurately reproduced as for the quasihole peaks below $F$. It is worth emphasizing that the emulator was constructed from snapshots at coupling strengths where both coupled cluster doubles (CCD)~\cite{Hagen2014} and the state-of-the-art self-consistent Green's function (SCGF) ADC(3)-D truncation~\cite{Barbieri::LectNotesPhys936, Marino2026} provide accurate results. Figure \ref{fig:SpecFunc_g1.3_partial_waves} demonstrates that the emulator retains remarkable accuracy when extrapolated to the strong-coupling regime at $g=1.3$, where CCD fails to converge and even SCGF calculations become inaccurate.

\begin{figure}
    \centering
    \includegraphics[width=\linewidth]{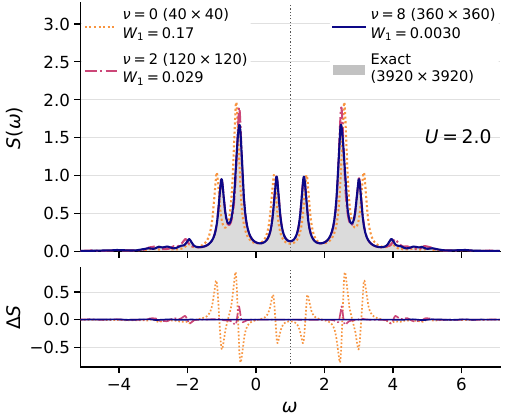}
    \caption{spectral-strength distribution of the Hubbard ring for the coupling $U=2.0$. The spectral peaks have been widened with a Lorentzian smearing with $\eta=0.1$, and results using $\nu=0, 2$, and $8$ for the emulator are presented. We also report in brackets the dimension of the emulator eigenvalue problem of equation~\eqref{eqECParticleSpace} for each $\nu$, together with the $W_1$ distance between exact and emulated spectral distributions. The bottom panel shows the difference between emulated and exact distribution. The Fermi energy separating holes and particles is shown as a vertical dotted line.}
    \label{fig:SpecFunc_orders_N8_L4_U2.00_t1.00_eta0.1}
\end{figure}

{\itshape Hubbard ring}---The Hamiltonian of the $N$-site Hubbard ring is
\begin{equation}
    H = -\sum_{\sigma=\uparrow,\downarrow}\sum_{j=1}^{N}
    \left(c^{\dagger}_{j\sigma}c_{j+1,\sigma}
    + \mathrm{h.c.}\right)
    + U\sum_{j=1}^{N}n_{j\uparrow}n_{j\downarrow},
\end{equation}
where $n_{j\sigma}=c^\dagger_{j\sigma}c_{j\sigma}$ and periodic boundary conditions are imposed through $c_{N+1,\sigma} \defeq c_{1\sigma}$. The hopping amplitude is taken as the unit of energy. We will consider the case of $N=8$ lattice sites at half-filling and choose snapshots at $U = \{-7.0, -4.0, 1.0, 3.0, 5.0\}$. Figure \ref{fig:SpecFunc_orders_N8_L4_U2.00_t1.00_eta0.1} shows the exact and emulated spectral-strength distribution of the Hubbard ring for $U=2.0$. Notice that by increasing the order $\nu$, we systematically increase the accuracy of the emulator. For $\nu = 8$, the emulator reproduces the full-order results, saving a factor of approximately $120$ in the number of matrix elements. Note that the results with $\nu = 0$, which require only $40 \times 40$ matrices, already show remarkably accurate results, especially for the large quasihole and quasiparticle peaks near the Fermi energy. 

\begin{figure*}
    \centering
    \includegraphics[width=0.7\linewidth]{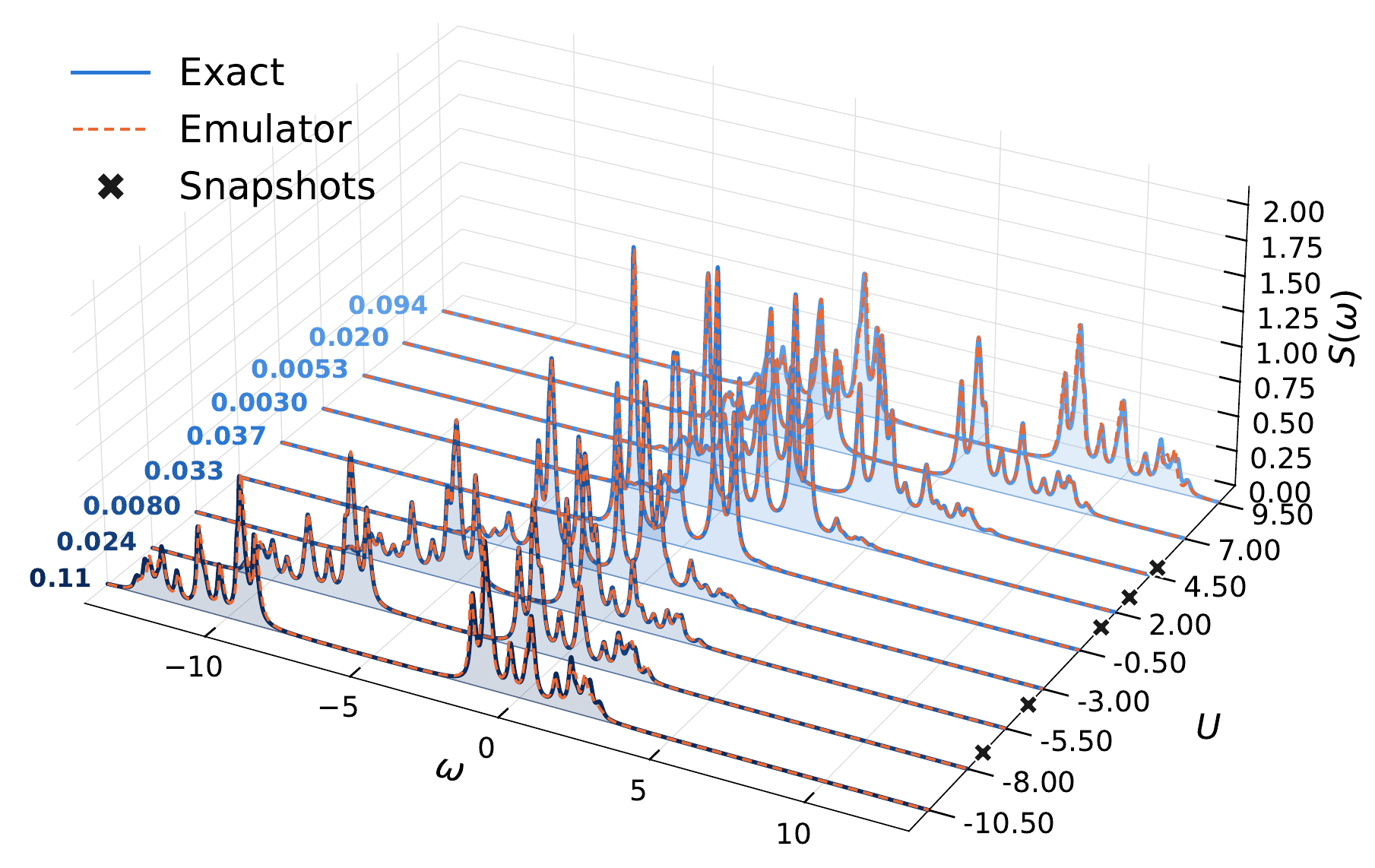}
    \caption{Exact and emulated spectral-strength distribution of the Hubbard ring shown for different values of the coupling $U$. The $W_1$ distance between emulated and exact distributions is shown on the left of each computed distribution.}
    \label{fig:SpecFunc3D_N8_L4_t1.00_eta0.1}
\end{figure*}

Figure \ref{fig:SpecFunc3D_N8_L4_t1.00_eta0.1} shows the exact and emulated spectral distributions of the Hubbard ring at different values of the coupling with $\nu = 8$. The emulator shows near-perfect agreement with the results of exact diagonalizations and small deviations (highlighted by the $W_1$ distance) only when moving far from the snapshots used in the emulator construction. On our laptop, computing the exact spectral distributions shown in Figure \ref{fig:SpecFunc3D_N8_L4_t1.00_eta0.1} takes 8.4 seconds, of which 4.9 seconds are devoted to the ground-state calculations. This involves diagonalizing a $4900 \times 4900$ matrix for the ground-state and two $3920 \times 3920$ matrices for the particle-addition and particle-removal sectors using Lanczos methods~\cite{Weikert1996}. In contrast, the ECGF requires the diagonalization of a $5 \times 5$ matrix for the ground-state at negligible computational cost, and two $360 \times 360$ matrices for the particle-addition and particle-removal sectors, for a total of just 1.3 seconds. The fixed time needed in the offline stage to obtain the emulator equations—including the calculations of the ground-state wavefunction at the snapshots—is 6.7 seconds. Further details on the computational cost of using exact diagonalization instead of Lanczos methods are provided in the Supplemental Material~\cite{SupplMat}. These results establish our emulation strategy as a promising approach for obtaining a reduced-order description of the Green's function capable of providing a substantial computational speedup and reduced memory requirements.

{\itshape Conclusion}---In this Letter, we have introduced a method for emulating the Green's function through reduced-order equations that preserve the parametric dependence of the underlying Hamiltonian, achieving substantial computational speedups and memory savings. This approach requires knowledge of the ground-state wave function, which prevents the emulator from being constructed directly from polynomially scaling SCGF calculations, where the many-body wavefunction is never explicitly computed. Nevertheless, we have rigorously established the parametric structure of the reduced-order matrices and derived upper bounds on their dimensions while systematically improving the description of spectral fragmentation through moment matching. These results provide a foundation for learning the matrix elements of the reduced-order eigenvalue problem directly from SCGF calculations, or even from experimental data, within the framework of parametric matrix models~\cite{Cook2025, cook2026}. Developing such an approach represents the natural next step toward extending the applicability of the present emulator.

{\itshape Acknowledgments}---This work used the DiRAC Data Intensive service (DIaL3) at the University of Leicester, managed by the University of Leicester Research Computing Service on behalf of the STFC DiRAC HPC Facility (www.dirac.ac.uk). The DiRAC service at Leicester was funded by BEIS, UKRI and STFC capital funding and STFC operations grants. DiRAC is part of the UKRI Digital Research Infrastructure.

{\itshape Data availability}---The data that support the findings of this article are not publicly available. The data are available from the authors upon reasonable request.

\nocite{Koltun1974}
\nocite{Migdal1958}
\bibliographystyle{apsrev4-2}
\bibliography{bibliography}

\end{document}


\begin{center}
\section*{Supplemental Material}
\end{center}

\renewcommand{\thefigure}{S\arabic{figure}}
\renewcommand{\thetable}{S\arabic{table}}
\renewcommand{\theequation}{S\arabic{equation}}
\renewcommand{\thepage}{\arabic{page}}
\setcounter{figure}{0}
\setcounter{table}{0}
\setcounter{equation}{0}
\setcounter{page}{1}

\newtheorem{theorem}{Theorem}
\newtheorem{corollary}{Corollary}[theorem]
\newtheorem{lemma}[theorem]{Lemma}

\subsection*{Construction of the eigenvector-continued Green's function}

In this section we present the full derivation of the emulator equations. To this end, consider a Hamiltonian with affine parameter dependence, 
\begin{equation} \label{eqFullHamiltonian}
    H(\theta) = H_0 + \theta H_1.
\end{equation} 
The ground-state Schr\"odinger equation for the $A$-particle system is
\begin{equation} \label{eqSchrodinger}
    H(\theta) \ket{\Psi_0^A(\theta)} = E_0^A(\theta) \ket{\Psi_0^A(\theta)},
\end{equation}
where $E_0^A(\theta)$ and $\ket{\Psi_0^A(\theta)}$ denote the ground-state energy and wavefunction, respectively.

We construct a reduced-order model using eigenvector continuation (EC)~\cite{Frame2018}. To this end, we solve the full eigenvalue problem~\eqref{eqSchrodinger} at a selected set of parameter values $\{\theta_i\}_{i=1}^{n_b}$, obtaining the ground-state wavefunctions $\{\ket{\Psi_0^A(\theta_i)}\}_{i=1}^{n_b}$, commonly referred to as \emph{snapshots}. Orthonormalizing the snapshots yields a basis $\{\ket{\psi_j}\}_{j=1}^{n_b}$ for the EC subspace. Projecting Eq.~\eqref{eqSchrodinger} onto this subspace leads to the reduced eigenvalue problem
\begin{equation} \label{eqREP_A}
    \left(\widetilde{H}_0^A + \theta \widetilde{H}_1^A\right)
    \ket{\widetilde{\Psi}_0^A(\theta)}
    = \widetilde{E}_0^A(\theta)
    \ket{\widetilde{\Psi}_0^A(\theta)},
\end{equation}
where a tilde denotes a reduced-order quantity and the superscript $A$ specifies the particle-number sector. Notice that the projected Hamiltonian retains the affine parameter dependence of the full Hamiltonian \eqref{eqFullHamiltonian}, and
\begin{equation}
\begin{split}
    \widetilde{H}^A_0
    &{}= \sum_{jk}\ket{\psi_j}
    \braket{\psi_j|H_0|\psi_k}
    \bra{\psi_k}, \\
    \widetilde{H}^A_1 &{}= \sum_{jk}\ket{\psi_j}
    \braket{\psi_j|H_1|\psi_k}
    \bra{\psi_k}. \\
\end{split}
\end{equation}

The approximate ground-state wavefunction is obtained by diagonalizing the projected Hamiltonian matrix
\begin{equation} \label{eqHRED}
    \left[\widetilde{H}^A(\theta)\right]_{jk}
    = \braket{\psi_j|H(\theta)|\psi_k},
\end{equation}
and takes the form
\begin{equation}
    \ket{\widetilde{\Psi}_0^A(\theta)}
    = \sum_{j=1}^{n_b}\beta_j(\theta)\ket{\psi_j},
\end{equation}
where $\beta_j(\theta)$ are the components of the eigenvector associated with the lowest eigenvalue $\widetilde{E}_0^A(\theta)$ of \eqref{eqHRED}. 

To emulate the Green's function, we use its Kallén-Lehmann representation~\cite{Kallen1952, Lehmann1954}, namely
\begin{equation} \label{eqLehmannSupplMat}
    G_{\alpha \beta} (\omega) = \sum_k \frac{\mathcal{Y}_\alpha^k \left(\mathcal{Y}_\beta^k\right)^*}{\omega -\epsilon_k^- - i\eta} + \sum_n \frac{\left(\mathcal{X}_\alpha^n\right)^*  \mathcal{X}_\beta^n}{\omega -\epsilon_n^+ + i\eta},
\end{equation}
where
\begin{equation} \label{eqCompsGFSupplMat}
\begin{split}
    \mathcal{X}_\alpha^n {}&\coloneqq \braket{\Psi_n^{A+1}|c_\alpha^\dagger|\Psi_0^{A}}; \\
    \mathcal{Y}_\alpha^k {}&\coloneqq \braket{\Psi_k^{A-1}|c_\alpha|\Psi_0^{A}}; \\
    \epsilon_n^+ {}&\coloneqq E_n^{A+1} - E_0^A; \\
    \epsilon_k^- {}&\coloneqq E_0^A - E_k^{A-1}.
\end{split}
\end{equation}
Here, $E_i^{A\pm1}$ and $\ket{\Psi_i^{A\pm1}}$ denote the energy and wavefunction of the $i$-th eigenstate of the $(A\pm1)$-particle system, respectively. We have omitted the explicit dependence on $\theta$ for brevity. The operators $c_\alpha$ and $c_\alpha^\dagger$ annihilate and create, respectively, a particle in the single-particle state $\alpha$ of the computational basis. To this end, we build subspaces of the $(A\pm1)$-particle systems starting from the states 
\begin{equation} \label{eqStartingParticleStates}
    \left\{ c_\alpha^\dagger \ket{\Psi_0^A(\theta_i)} \right\}_{i=1, \ldots, n_b;\ \alpha = 1, \ldots, D}
\end{equation} and 
\begin{equation} \label{eqStartingHoleStates}
    \left\{ c_\alpha \ket{\Psi_0^A(\theta_i)} \right\}_{i=1, \ldots, n_b;\ \alpha = 1, \ldots, D},
\end{equation} where $D$ is the dimension of the computational basis. These subspaces can be systematically enriched in two distinct ways.

\begin{itemize}
    \item \emph{Power grouping method}. We first consider the particle sector of the Green's function and construct the states obtained by applying the Hamiltonian $\mu$ times to the starting vectors in Eq.~\eqref{eqStartingParticleStates}:
    \begin{equation} \label{eqStatesMoments}
        \left(H_0+\theta H_1\right)^\mu c_\alpha^\dagger \ket{\Psi_0^A(\theta_i)}.
    \end{equation}
    Including these states in the reduced subspace allows the emulator to capture higher-order moments of the spectral-strength distribution, as demonstrated in the next section. However, this construction introduces an explicit dependence of the subspace on the parameter $\theta$ through the Hamiltonian $H(\theta)$. To eliminate this dependence, we expand the powers of the Hamiltonian as
    \begin{equation}
        \left(H_0+\theta H_1\right)^\mu
        = \sum_{k=0}^\mu \theta^k O_k^{(\mu)},
    \end{equation}
    where $O_k^{(\mu)}$ denotes the sum of all ordered products containing $k$ factors of $H_1$ and $\mu-k$ factors of $H_0$. The operators $O_k^{(\mu)}$ are therefore independent of $\theta$.
    
    We then construct the subspace
    \begin{equation} \label{eqStatesPowerGrouping}
        \mathrm{span}\left\{
        O_k^{(\mu)} c_\alpha^\dagger \ket{\Psi_0^A(\theta_i)}
        \right\}_{k=0, \ldots, \mu;\ \mu=0,\ldots,\nu;\ i=1,\ldots,n_b;\ \alpha=1,\ldots,D}.
    \end{equation}
    This subspace is generated by $\tfrac{1}{2} D n_b(\nu+1)(\nu+2)$ vectors and contains all states of the form~\eqref{eqStatesMoments} for $\mu\leq\nu$, since they can be expressed as a linear combination of the vectors in Eq.~\eqref{eqStatesPowerGrouping}. Orthonormalizing these vectors and discarding linearly dependent components, we obtain the orthonormal basis
    \begin{equation}
        \left\{\ket{b_j^{A+1}}\right\},
    \end{equation}
    whose dimension is at most $\tfrac{1}{2} D n_b(\nu+1)(\nu+2)$.
    
    Similarly, the corresponding subspace for the $(A-1)$-particle sector is constructed starting from
    \begin{equation} \label{eqStatesPowerGroupingHoles}
        \mathrm{span}\left\{
        O_k^{(\mu)} c_\alpha \ket{\Psi_0^A(\theta_i)}
        \right\}_{k=0,\ldots,\mu;\ \mu=0,\ldots,\nu;\ i=1,\ldots,n_b;\ \alpha=1,\ldots,D},
    \end{equation}
    which yields the orthonormal basis
    \begin{equation}
        \left\{\ket{b_j^{A-1}}\right\}.
    \end{equation}
    The power grouping method thus produces reduced subspaces whose dimensions grow at most quadratically with $\nu$. As shown in the next section, this construction guarantees that the emulator reproduces the moments of the spectral-strength distribution up to order $2\nu+1$, within the accuracy of the EC approximation to the ground state.
    
    \item \emph{Diagonal approximation technique}. We introduce an alternative construction of the $(A\pm1)$-particle subspaces that requires a basis size growing at most linearly with $\nu$, while guaranteeing exact moment matching at the snapshot parameter values.

    Focusing again on the particle sector of the Green's function, we construct the subspace
    \begin{equation}
        \mathrm{span}\left\{
        H^\mu(\theta_i)c_\alpha^\dagger\ket{\Psi_0^A(\theta_i)}
        \right\}_{\mu=0,\ldots,\nu;\ i=1,\ldots,n_b;\ \alpha=1,\ldots,D}.
    \end{equation}
    Orthonormalizing the spanning vectors and discarding linearly dependent components, we obtain the orthonormal basis
    \begin{equation}
        \left\{\ket{b_j^{A+1}}\right\}.
    \end{equation}
    Analogously, for the hole sector, we construct the subspace
    \begin{equation}
        \mathrm{span}\left\{
        H^\mu(\theta_i)c_\alpha\ket{\Psi_0^A(\theta_i)}
        \right\}_{\mu=0,\ldots,\nu;\ i=1,\ldots,n_b;\ \alpha=1,\ldots,D},
    \end{equation}
    and obtain the orthonormal basis
    \begin{equation}
        \left\{\ket{b_j^{A-1}}\right\}.
    \end{equation}
    In both particle-number sectors, the resulting subspace has a dimension of at most $D n_b(\nu+1)$, scaling linearly with $\nu$, in contrast to the quadratic scaling of the power grouping method.
\end{itemize}

In the following, we omit the superscript $A\pm1$ from the basis states $\ket{b_j^{A \pm 1}}$ whenever the particle-number sector is clear from the context.
We now turn to the construction of the emulated Green's function, starting with the $(A+1)$-particle sector. Once an orthonormal basis $\{\ket{b_j^{A+1}}\}$ has been constructed using either of the two techniques described above, we define the projected Hamiltonian and coupling matrix as
\begin{equation} \label{eqHAp1Red}
    \left[\widetilde{H}^{A+1}(\theta)\right]_{jk}
    \defeq \braket{b_j|H(\theta)|b_k},
\end{equation}
and
\begin{equation}
    \left[\widetilde{N}^{A+1}\right]_{k,j\alpha}
    \defeq \braket{b_k|c_\alpha^\dagger|\psi_j},
\end{equation}
respectively. While the coupling matrix $\widetilde{N}^{A+1}$ is independent of $\theta$, the projected Hamiltonian retains the affine parameter dependence of the original Hamiltonian,
\begin{equation} \label{eqEPAp1}
    \widetilde{H}^{A+1}(\theta)
    = \widetilde{H}_0^{A+1}+\theta\widetilde{H}_1^{A+1},
\end{equation}
where
\begin{equation}
    \left[\widetilde{H}_{r}^{A+1}\right]_{jk}
    \defeq \braket{b_j|H_r|b_k},
    \qquad r=0,1.
\end{equation}

Diagonalizing the projected Hamiltonian \eqref{eqHAp1Red} yields the approximate eigenvalues $\widetilde{E}_n^{A+1}(\theta)$ and the corresponding eigenstates
\begin{equation}
    \ket{\widetilde{\Psi}_n^{A+1}(\theta)}
    = \sum_j A_j^{n,+}(\theta)\ket{b_j},
\end{equation}
where $A_j^{n,+}(\theta)$ are the components (labelled by $j$) of the eigenvectors (labelled by $n$) of $\widetilde{H}^{A+1}(\theta)$.

The poles and spectroscopic amplitudes of the particle Green's function are then obtained from Eq.~\eqref{eqCompsGFSupplMat} as
\begin{equation}
    \widetilde{\epsilon}_n^+
    \defeq \widetilde{E}_n^{A+1}-\widetilde{E}_0^A,
\end{equation}
and
\begin{equation}
    \widetilde{\mathcal{X}}_\alpha^n \defeq \braket{\widetilde{\Psi}_n^{A+1}|c_\alpha^\dagger|\widetilde{\Psi}_0^A} = \sum_{jl}
    \left(A_l^{n,+}\right)^* \beta_j \braket{b_l|c_\alpha^\dagger|\psi_j} = \sum_{jl} \left(A_l^{n,+}\right)^* \widetilde{N}^{A+1}_{l,j\alpha} \beta_j.
\end{equation}
Here and in the following, we suppress the explicit dependence on $\theta$ when no ambiguity arises.

The $(A-1)$-particle sector is constructed analogously. Given the corresponding orthonormal basis $\{\ket{b_j^{A-1}}\}$, we define
\begin{equation} \label{eqHAm1Red}
    \left[\widetilde{H}^{A-1}(\theta)\right]_{jk}
    \defeq \braket{b_j^{A-1}|H(\theta)|b_k^{A-1}},
\end{equation}
and
\begin{equation}
    \left[\widetilde{N}^{A-1}\right]_{k,j\alpha}
    \defeq \braket{b_k^{A-1}|c_\alpha|\psi_j}.
\end{equation}
As in the particle sector, the projected Hamiltonian depends affinely on $\theta$, while the coupling matrix is parameter independent. Diagonalizing $\widetilde{H}^{A-1}(\theta)$ yields the approximate eigenvalues $\widetilde{E}_k^{A-1}(\theta)$ and eigenstates
\begin{equation}
    \ket{\widetilde{\Psi}_k^{A-1}(\theta)}
    = \sum_j A_j^{k,-}(\theta)\ket{b_j^{A-1}}.
\end{equation}
The corresponding poles and spectroscopic amplitudes of the hole Green's function are
\begin{equation}
    \widetilde{\epsilon}_k^-
    \defeq \widetilde{E}_0^A-\widetilde{E}_k^{A-1},
\end{equation}
and
\begin{equation}
    \widetilde{\mathcal{Y}}_\alpha^k
    \defeq \braket{\widetilde{\Psi}_k^{A-1}|c_\alpha|\widetilde{\Psi}_0^A} = \sum_{jl}
    \left(A_l^{k,-}\right)^* \widetilde{N}^{A-1}_{l,j\alpha} \beta_j.
\end{equation}

Finally, the approximation to the single-particle Green's function, which we refer to as the eigenvector-continued Green's function (ECGF), takes the form
\begin{equation} 
    \widetilde{G}_{\alpha \beta} (\omega) = \sum_k \frac{\widetilde{\mathcal{Y}}_\alpha^k \left(\widetilde{\mathcal{Y}}_\beta^k\right)^*}{\omega -\widetilde{\epsilon}_k^- - i\eta} + \sum_n \frac{\left(\widetilde{\mathcal{X}}_\alpha^n\right)^*  \widetilde{\mathcal{X}}_\beta^n}{\omega -\widetilde{\epsilon}_n^+ + i\eta}.
\end{equation}

\subsection{Properties of the emulator} 

In this section, we show the properties of the ECGF, and in particular its ability of capturing the density matrix, high-order moments of the spectral-strength distribution, and large spectroscopic peaks. Consider the general case of a value of $\theta$ not used for the snapshots. 

We start with properties that are shared by both methods of constructing the $A\pm1$ subspaces, namely, the power grouping technique and the diagonal approximation.

We start with a lemma that will be widely employed in the following theorems, namely

\begin{lemma} \label{Lemma1}
    Let $\ket{\widetilde{\Psi}_i^{A\pm1}}$ be an eigenstate of the projected Hamiltonian $\widetilde{H}^{A\pm1}$ with eigenvalue $\widetilde{E}_i^{A\pm1}$. Then, for any state $\ket{v}$ in the subspace spanned by $\{\ket{b_j^{A\pm1}}\}$,
    \begin{equation}
        \braket{v|H|\widetilde{\Psi}_i^{A\pm1}}
        = \widetilde{E}_i^{A\pm1}
        \braket{v|\widetilde{\Psi}_i^{A\pm1}}.
    \end{equation}

    \begin{proof}
        We prove the result for the particle sector; the hole sector follows analogously. Since both $\ket{v}$ and $\ket{\widetilde{\Psi}_n^{A+1}}$ belong to the subspace spanned by $\{\ket{b_j}\}$, we have
        \begin{equation}
            \braket{v|H|\widetilde{\Psi}_n^{A+1}} = \sum_{jk} \braket{v|b_j} \braket{b_j|H|b_k} \braket{b_k|\widetilde{\Psi}_n^{A+1}} = \braket{v|\widetilde{H}^{A+1}|\widetilde{\Psi}_n^{A+1}} = \widetilde{E}_n^{A+1} \braket{v|\widetilde{\Psi}_n^{A+1}},
        \end{equation}
        where the last equality follows from the reduced eigenvalue equation in the $(A+1)$-particle sector. A similar proof applies to the $(A-1)$-particle sector.
    \end{proof}
\end{lemma}

\begin{theorem} \label{ThDensityMatrix}
    The emulator captures the one-body density matrix up to the approximation given by eigenvector continuation for all $\nu \geq 0$.
    \begin{proof}
        The exact one-body density matrix is defined as
        \begin{equation}
            \rho_{\alpha \beta} \defeq \braket{\Psi_0^A|c_\beta^\dagger c_\alpha|\Psi_0^A}.
        \end{equation}
        
        Computing the one-body density matrix from the emulated Green's function, we obtain
        \begin{equation} \label{eqEmulRho}
            \widetilde{\rho}_{\alpha \beta} = \sum_k \left( \widetilde{\mathcal{Y}}^k_{\beta} \right)^* \widetilde{\mathcal{Y}}^k_{\alpha} = \sum_k \braket{\widetilde{\Psi}_0^A|c_\beta^\dagger|\widetilde{\Psi}_k^{A-1}} \braket{\widetilde{\Psi}_k^{A-1}|c_\alpha|\widetilde{\Psi}_0^A} = \braket{\widetilde{\Psi}_0^A|c_\beta^\dagger c_\alpha|\widetilde{\Psi}_0^{A}},
        \end{equation}
        where we have used the fact that $c_\alpha \ket{\widetilde{\Psi}_0^{A}}$ is contained in the space spanned by the eigenstates $\ket{\widetilde{\Psi}_k^{A-1}}$.
        Equation \eqref{eqEmulRho} is the exact one-body density matrix up to the EC approximation to the ground-state.
    \end{proof}
\end{theorem}

\begin{theorem} 
    Poles with large spectroscopic factors are accurately captured by the emulator. In particular, the spectroscopic factor provides a lower bound on the fidelity with which that eigenstate can be represented by the eigenvector-continued Green's function.
    \begin{proof}
        We define the spectral amplitude
        \begin{equation}
            \overline{\mathcal{X}}^n_\alpha \coloneqq \braket{\Psi_n^{A+1}|c_\alpha^\dagger|\widetilde{\Psi}_0^A}
        \end{equation}
        and build a representation of the state $\ket{\Psi_n^{A+1}}$ in the space spanned by the emulator basis as
        \begin{equation} \label{eq_phi_n_ApluOne}
            \ket{\phi_n^{A+1}} \coloneqq \frac{1}{\sqrt{Z_n}} \sum_{\alpha} \left( \overline{\mathcal{X}}^n_\alpha\right)^*  c_\alpha^\dagger \ket{\widetilde{\Psi}_0^A},
        \end{equation}
        where
        \begin{equation}
            Z_n \defeq \sum_{\alpha, \delta} \overline{\mathcal{X}}^n_\delta \braket{\widetilde{\Psi}_0^A|c_\delta c_\alpha^\dagger|\widetilde{\Psi}_0^A} \left( \overline{\mathcal{X}}^n_{\alpha} \right)^*
        \end{equation}
        is the normalization factor. The overlap between this state and the exact eigenstate is given by
        \begin{equation}
            \braket{\Psi_n^{A+1}|\phi_n^{A+1}} = \frac{1}{\sqrt{Z_n}} \sum_{\alpha} \left( \overline{\mathcal{X}}^n_{\alpha} \right)^* \braket{\Psi_n^{A+1}|c_\alpha^\dagger|\widetilde{\Psi}_0^A} = \frac{1}{\sqrt{Z_n}} \sum_{\alpha} \left\lvert \overline{\mathcal{X}}^n_{\alpha} \right\rvert^2 
            \frac{1}{\sqrt{Z_n}} \overline{SF}^+_n,
        \end{equation}
        where we have defined $\overline{SF}^+_n \coloneqq \sum_{\alpha} \left\lvert \overline{\mathcal{X}}^n_{\alpha} \right\rvert^2$.
        Since all eigenvalues of $\braket{\widetilde{\Psi}_0^A|c_\delta c_\alpha^\dagger|\widetilde{\Psi}_0^A}$ are smaller than or equal to one, we have 
        \begin{equation}
            Z_n = \sum_{\alpha \delta} \overline{\mathcal{X}}^n_\delta \braket{\widetilde{\Psi}_0^A|c_\delta c_\alpha^\dagger|\widetilde{\Psi}_0^A} \left( \overline{\mathcal{X}}^n_{\alpha} \right)^* \leq \overline{SF}^+_n.
        \end{equation}
        Therefore,
        \begin{equation} \label{eqBoundEumlator}
            \left\lvert \braket{\Psi_n^{A+1}|\phi_n^{A+1}} \right\rvert^2 = \frac{\left(\overline{SF}_n^+\right)^2}{Z_n} \geq \overline{SF}_n^+.
        \end{equation}
    \end{proof}
\end{theorem}

Notice that this proof relies on building a representative state in the $(A+1)$-particle subspace as $\ket{\phi_n^{A+1}} \propto \sum_{\alpha} \left( \overline{\mathcal{X}}^n_\alpha\right)^*  c_\alpha^\dagger \ket{\widetilde{\Psi}_0^A}$. In general, we can access a larger space; even in the case of $\nu=0$, we can form states of the form $\ket{\phi_n^{A+1}} \propto \sum_{j\alpha}  A_{j \alpha}^n c^{\dagger}_\alpha \ket{\psi_j}$. Hence, in practice, this bound is often improved.

\begin{theorem} \label{Th_Moment_zero}
    The emulator reproduces the exact number of holes and particles, given by the $0$-th moment of the spectral strength distribution, for all $\nu \geq 0$.
    \begin{proof}
        The $0$-th moment of the emulated hole spectral-strength distribution is
        \begin{equation}
            \widetilde{m}^{(0)}_{-} = \sum_\alpha \sum_{k} \left\lvert \widetilde{\mathcal{Y}}^{k}_\alpha \right\rvert^2 = \sum_{jn} \sum_\alpha \sum_k \beta_n^* \braket{\psi_n| c_\alpha^\dagger| \widetilde{\Psi}_{k}^{A-1}} \braket{\widetilde{\Psi}_{k}^{A-1}|c_\alpha|\psi_j} \beta_{j} = \sum_{jn} \beta_n^* \braket{\psi_n|\sum_\alpha c_\alpha^\dagger c_\alpha|\psi_j} \beta_{j} = A,
        \end{equation}
        where we have used the fact that $c_\alpha^\dagger\ket{\widetilde{\Psi}_{k}^{A-1}}$ lies in the subspace spanned by $\{ \ket{\widetilde{\Psi}_k^{A-1}} \}_k$. 
        Similarly, we find
        \begin{equation}
            \widetilde{m}^{(0)}_{+} = D - A.
        \end{equation}
    \end{proof}
\end{theorem}

\begin{theorem} \label{ThFirstMoment}
    The emulator reproduces the first moment of the hole and particle spectral-strength distributions.
    \begin{proof}
        The first moment of the exact hole-part of the spectral strength distribution is 
        \begin{equation} \label{EqExactMomentOne}
        \begin{split}
            m_-^{(1)} &{}= \sum_k \epsilon_k^- \sum_\alpha \left\lvert \mathcal{Y}^{k}_\alpha \right\rvert^2 = \sum_k \left( E_0^A - E_k^{A-1} \right) \sum_\alpha \left\lvert \mathcal{Y}^{k}_\alpha \right\rvert^2 = \sum_k \left( E_0^A - E_k^{A-1} \right) \sum_\alpha \braket{\Psi_0^A|c_\alpha^\dagger|\Psi_k^{A-1}} \braket{\Psi_k^{A-1}|c_\alpha|\Psi_0^A} \\
            &{}= E_0^A A - \braket{\Psi_0^A|c_\alpha^\dagger H c_\alpha|\Psi_0^A}.
        \end{split}
        \end{equation}
        The first moment of the emulated hole-part of the spectral-strength distribution is 
        \begin{equation} \label{eqEmulatorMomentOne1}
            \widetilde{m}_-^{(1)} = \sum_k \widetilde{\epsilon}_k^- \sum_\alpha \left\lvert \widetilde{\mathcal{Y}}^{k}_\alpha \right\rvert^2 = \sum_k \left( \widetilde{E}_0^A - \widetilde{E}_k^{A-1} \right) \sum_\alpha \left\lvert \widetilde{\mathcal{Y}}^{k}_\alpha \right\rvert^2 = A \widetilde{E}_0^A - \sum_k \widetilde{E}_k^{A-1}  \sum_\alpha \left\lvert \widetilde{\mathcal{Y}}^{k}_\alpha \right\rvert^2,
        \end{equation}
        where we have used Theorem~\ref{Th_Moment_zero}.
        Let us consider the second term of the last line of this equation
        \begin{equation} \label{eqEmulatorMomentOne2}
        \begin{split}
            \sum_k \widetilde{E}_k^{A-1}  \sum_\alpha \left\lvert \widetilde{\mathcal{Y}}^{k}_\alpha \right\rvert^2 &{}= \sum_k \widetilde{E}_k^{A-1}  \braket{\widetilde{\Psi}_0^A|c_\alpha^\dagger|\widetilde{\Psi}_k^{A-1}} \braket{\widetilde{\Psi}_k^{A-1}|c_\alpha|\widetilde{\Psi}_0^A} \\
            &{}= \sum_k \widetilde{E}_k^{A-1}\sum_{jl} \beta_j^* \beta_l   \braket{\psi_j|c_\alpha^\dagger|\widetilde{\Psi}_k^{A-1}} \braket{\widetilde{\Psi}_k^{A-1}|c_\alpha|\psi_l} \\
            &{}= \sum_k \sum_{jl} \beta_j^* \beta_l \braket{\psi_j|c_\alpha^\dagger H|\widetilde{\Psi}_k^{A-1}} \braket{\widetilde{\Psi}_k^{A-1}|c_\alpha|\psi_l} \\
            &{}= \sum_{jl} \beta_j^* \beta_l \braket{\psi_j|c_\alpha^\dagger H c_\alpha|\psi_l} = \braket{\widetilde{\Psi}_0^A|c_\alpha^\dagger H c_\alpha|\widetilde{\Psi}_0^A},
        \end{split}
        \end{equation}
        where we have used Lemma\ref{Lemma1} and the fact that $c_\alpha \ket{\psi_j}$ lies in the space spanned by $\ket{\widetilde{\Psi}_k^{A-1}}$.
        Comparing equations \eqref{eqEmulatorMomentOne1} and \eqref{eqEmulatorMomentOne2} with equation \eqref{EqExactMomentOne} proves that the emulator captures the first moment of the hole-sector of the spectral-strength distribution up to the approximation set by EC. An analogous proof applies to the particle sector of the spectral-strength distribution.
    \end{proof}
\end{theorem}

\begin{corollary} \label{CorGMKEnergy}
    The energy of the eigenvector-continued Green's function calculated with the Galitskii-Migdal-Koltun sum rule gives the same result as eigenvector continuation.

    \begin{proof}
        The Galitskii-Migdal-Koltun sum rule~\cite{Koltun1974, Migdal1958} is 
        \begin{equation} \label{eqGMK_corollary}
            E_0^{\mathrm{GMK}} = \frac{1}{2} \sum_{k} \sum_{\alpha \beta} \left( T_{\alpha \beta} + \epsilon_k^- \delta_{\alpha \beta} \right) \mathcal{Y}_\beta^k (\mathcal{Y}_\alpha^k)^* = \frac{1}{2} \sum_{\alpha \beta} T_{\alpha \beta} \rho_{\beta \alpha} + \frac{1}{2} m_-^{h}.
        \end{equation}
        By Theorems \ref{ThDensityMatrix} and \ref{ThFirstMoment}, the emulator reproduces both terms of the final expression in \eqref{eqGMK_corollary} up to the EC approximation; hence, the ground-state energy is reproduced up to the EC approximation.
    \end{proof}
\end{corollary}

\begin{theorem} \label{ThMoments}
    In the power grouping approach, moments up to order $2 \nu +1$ of the particle and the hole sectors of the spectral-strength distribution are reproduced exactly, up to the approximation to the ground-state given by eigenvector continuation.
    \begin{proof}
        The $\mu$-th moment of the exact particle spectral strength distribution is defined as
        \begin{equation} \label{eq_moments_mu}
            m^{(\mu)}_+ \defeq \sum_n \left(\epsilon_n^+ \right)^\mu \sum_\alpha \left\lvert \mathcal{X}_\alpha^n \right\rvert^2 = \sum_n \left(E_n^{A+1} - E_0^A\right)^\mu \sum_\alpha \left\lvert \mathcal{X}_\alpha^n \right\rvert^2.
        \end{equation}
        We also introduce the reduced moments
        \begin{equation}
            m^{(\mu), \, \mathrm{red}}_+ \defeq \sum_n \left(E_n^{A+1}\right)^\mu \sum_\alpha \left\lvert \mathcal{X}_\alpha^n \right\rvert^2.
        \end{equation}
        Since Corollary~\eqref{CorGMKEnergy} guarantees that the ground-state energy is reproduced within the accuracy of the EC approximation, a binomial expansion of Eq.~\eqref{eq_moments_mu} shows that reproducing the reduced moments up to order $2\nu+1$ within the accuracy of the EC approximation also guarantees that the moments of the spectral-strength distribution are reproduced to the same order $2\nu + 1$ and EC accuracy. Therefore, in the following, we focus on the reduced moments.

        The first reduced moment of the emulated particle spectral strength distribution is 
        \begin{equation} \label{eqExactMoment}
                m^{(\mu), \, \mathrm{red}}_{+} = \sum_n \left( E_n^{A+1} \right)^\mu  \sum_\alpha \left\lvert \mathcal{X}^{n}_\alpha \right\rvert^2 = \sum_n \left( E_n^{A+1} \right)^\mu \sum_\alpha \braket{\Psi_0^A|c_\alpha|\Psi_n^{A+1}} \braket{\Psi_n^{A+1}|c_\alpha^\dagger|\Psi_0^A} = \sum_\alpha \braket{\Psi_0^{A}|c_\alpha H c_\alpha^\dagger|\Psi_0^A}.
        \end{equation}

        We now consider the emulated spectral function. Applying $H^{\mu}(\theta)$, with $\mu \leq \nu + 1$, to the state $\ket{\widetilde{\Psi}_n^{A+1}}$ and computing the product with $\bra{\psi_l} c_\beta$, we obtain
        \begin{equation}
            \braket{\psi_l| c_\beta H^{\mu} |\widetilde{\Psi}_n^{A+1}} = \sum_k \braket{\psi_l | c_\beta H^{\mu - 1}|b_k} \braket{b_k|H|\widetilde{\Psi}_n^{A+1}} = \widetilde{E}_{n}^{A+1} \braket{\psi_l| c_\beta H^{\mu - 1} |\widetilde{\Psi}_n^{A+1}} =  \left( \widetilde{E}_{n}^{A+1} \right)^\mu \braket{\psi_l| c_\beta |\widetilde{\Psi}_n^{A+1}},
        \end{equation}
        where we have used Lemma \ref{Lemma1} the fact that $H^{\mu - 1} c_\beta^\dagger \ket{\psi_l}$ lies in the space spanned by $\{ \ket{b_k} \}$ for all $\mu \leq \nu + 1$.

        If we define $\zeta \leq \nu + 1$, $\sigma \leq \nu$ such that  $\zeta + \sigma = \mu \leq 2 \nu + 1$, we have
        
        \begin{equation}
            \begin{split} \label{eqRedMoment}
                \widetilde{m}^{(\mu), \, \mathrm{red}}_{+} &{}= \sum_n \left( \widetilde{E}_n^{A+1} \right)^\mu  \sum_\alpha \left\lvert \widetilde{\mathcal{X}}^{n}_\alpha \right\rvert^2 = \sum_n \left( \widetilde{E}_n^{A+1} \right)^{\zeta + \sigma}  \sum_\alpha \sum_{km} \beta^*_k \beta_m  \braket{\psi_k |c_\alpha|\widetilde{\Psi}_n^{A+1}} \braket{\widetilde{\Psi}_n^{A+1}| c_\alpha^\dagger |\psi_m} \\
                &{}= \sum_n \sum_{\alpha} \sum_{km} \beta^*_k \beta_m  \braket{\psi_k |c_\alpha H^{\zeta}|\widetilde{\Psi}_n^{A+1}} \braket{\widetilde{\Psi}_n^{A+1}| H^{\sigma}c_\alpha^\dagger |\psi_m} = \sum_{\alpha} \sum_{km} \beta^*_k \beta_m \braket{\psi_k |c_\alpha H^{\mu} c_\alpha^\dagger|\psi_m} \\
                &{}= \sum_\alpha \braket{\widetilde{\Psi}_0^A |c_\alpha H^{\mu} c_\alpha^\dagger|\widetilde{\Psi}_0^A},
            \end{split}
        \end{equation}
    where we have used the fact that the vectors $\left\{ \ket{\widetilde{\Psi}_n^{A+1}}\right\}_n$ span the full $\left\{\ket{b_l} \right\}_l$ subspace containing all states of the form $H^\sigma c_\alpha^\dagger \ket{\psi_m}$, with $\sigma \leq \nu$.
    The last line of \eqref{eqRedMoment} is the exact moment of the spectral-strength distribution in equation \eqref{eqExactMoment} up to the EC approximation to the ground-state.
    \end{proof}
\end{theorem}

Notice that this implies that for the emulator in the power grouping approximation, we must have

\begin{corollary} 
    In the power grouping approach, moments up to order $2 \nu +1$ of the particle and the hole sectors of the spectral-strength distribution are reproduced exactly at the location of the snapshots.

    \begin{proof}
        Since eigenvector continuation is exact at the snapshots, by Theorem \ref{ThMoments}, the moments are exactly reproduced at the snapshots. 
    \end{proof}
\end{corollary}

This result is weaker than Theorem~\ref{ThMoments}, which bounds the error at parameter values $\theta$ outside the snapshot set in terms of the accuracy of the EC approximation to the ground state. Here, no such connection is established, and the error may include contributions beyond those arising from the EC approximation. 

This weaker form is precisely what is enforced by the diagonal approximation technique.

\begin{theorem}
    In the diagonal approximation approach, moments up to order $2 \nu +1$ of the particle and hole sectors of the spectral-strength distribution are reproduced exactly at the location of the snapshots.
    \begin{proof}
        The proof of this Theorem follows the proof of Theorem~\ref{ThMoments} by considering that at the snapshots, $\left\{ \sum_l A^n_l \ket{b_l} \right\}_n$ span the subspace containing $H^\sigma c_\alpha^\dagger \ket{\psi_m}$ for all $\sigma \leq \nu$.
    \end{proof}
\end{theorem}

\subsection{Wasserstein-1 distance}
We recall the properties of the Wasserstein-1 distance and how it can be employed to represent the accuracy of the emulator.
Let $(X,\rho)$ be a metric space, where $X$ is a set and $\rho$ is a distance function. Let $f$ and $g$ be two probability distributions on $X$. The Wasserstein-1 distance between $f$ and $g$ is defined as
\begin{equation} \label{eqWassersteinGeneral}
    W_1(f,g) \defeq \inf_{\pi \in \Pi(f,g)} \mathbb{E}_{(x,y)\sim\pi}\left[\rho(x,y)\right],
\end{equation}
where $\Pi(f,g)$ denotes the set of all joint probability distributions $\pi(x,y)$ whose marginals are $f(x)$ and $g(y)$, and $\mathbb{E}_{(x,y)\sim\pi}[\cdot]$ denotes the expectation value with respect to $\pi$. The Wasserstein-1 distance thus quantifies the minimum cost of transporting probability mass from one distribution to the other, where the cost of moving a unit of mass from $x$ to $y$ is given by $\rho(x,y)$~\cite{Clark1984}. For this reason, it is also known as the \emph{earth mover's distance}~\cite{peyre2020}. 

We employ the Wasserstein-1 distance to quantify the difference between the exact and emulated spectral-strength distributions. Theorem~\ref{Th_Moment_zero} guarantees that both distributions have the same normalization, allowing for a direct comparison. We first normalize them to unity,
\begin{equation}
    \overline{S}_x(\omega) = \frac{S_x(\omega)}{D},
    \qquad x=\mathrm{exact},\mathrm{emul}.
\end{equation}
Since the spectral-strength distributions are defined over the real energy axis, we adopt the standard Euclidean distance
\begin{equation}
    \rho(\omega_1,\omega_2) \defeq |\omega_2-\omega_1|.
\end{equation}
In this case, the Wasserstein-1 distance can be expressed in terms of the cumulative spectral-strength distributions~\cite{Vallender1974},
\begin{equation}
    W_1 = \int_{-\infty}^{+\infty} d\omega \,
    \left|
    \overline{CS}_{\mathrm{emul}}(\omega)
    - \overline{CS}_{\mathrm{exact}}(\omega)
    \right|,
\end{equation}
where
\begin{equation}
    \overline{CS}_x(\omega)
    \defeq \int_{-\infty}^{\omega} d\omega'\,
    \overline{S}_x(\omega'),
    \qquad x=\mathrm{exact},\mathrm{emul}.
\end{equation}

To illustrate the physical meaning of this distance, consider two normalized spectral-strength distributions,
\begin{equation}
    \overline{S}_1(\omega) = \delta(\omega-a),
\end{equation}
and
\begin{equation}
    \overline{S}_2(\omega)
    = \frac{1}{2}\delta(\omega-b)
    + \frac{1}{2}\delta(\omega-c),
\end{equation}
with $a<b<c$. Their cumulative distributions are
\begin{equation}
    \overline{CS}_1(\omega) = \theta(\omega-a),
\end{equation}
and
\begin{equation}
    \overline{CS}_2(\omega)
    = \frac{1}{2}\theta(\omega-b)
    + \frac{1}{2}\theta(\omega-c),
\end{equation}
where $\theta$ denotes the Heaviside step function. Evaluating the Wasserstein-1 distance gives
\begin{equation}
    W_1 = \frac{1}{2}(b-a) + \frac{1}{2}(c-a).
\end{equation}
This result has a simple interpretation in terms of spectral strength redistribution: transforming $\overline{S}_1$ into $\overline{S}_2$ requires splitting the peak at $\omega=a$ into two peaks of equal strength and transporting half of the spectral strength over a distance $b-a$ and the other half over a distance $c-a$. The Wasserstein-1 distance is the total transport cost, obtained by weighting each displacement by the amount of spectral strength transported.

\subsection{Comparison between power grouping and diagonal approximation technique}

\begin{figure}
    \centering
    \includegraphics[width=0.75\linewidth]{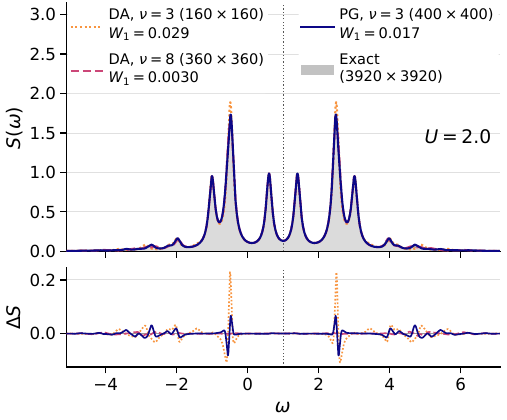}
    \caption{Comparison between the spectral-strength distribution at the coupling $U=2.0$ computed with the power-grouping (PG) and diagonal approximation (DA) methods. The spectral-strength distributions have been smeared with a Lorentzian of width $\eta = 0.1$. We also report in brackets the dimension of the emulator eigenvalue problem of equation \eqref{eqEPAp1} for each $\nu$, together with the $W_1$ distance between exact and emulated spectral distributions. The bottom panel shows the difference between emulated and exact distribution. The Fermi energy separating holes and particles is shown as a vertical dotted line.}
    \label{fig:PGvDA}
\end{figure}

Figure~\ref{fig:PGvDA} compares the spectral-strength distributions obtained with the diagonal approximation (DA) and power grouping (PG) methods for the Hubbard ring, with a coupling $U=2.0$. The DA method consistently provides a more accurate representation of the spectral-strength distribution at a lower computational cost. In particular, the DA method with $\nu=3$ achieves an accuracy comparable to that of the PG method, as quantified by the Wasserstein-1 distance $W_1$, while requiring substantially smaller matrices. Increasing the DA order to $\nu=8$ further improves the agreement with the exact spectral-strength distribution, yielding significantly smaller Wasserstein-1 distances than the PG method with matrices approximately $10\%$ smaller.

These results highlight the computational efficiency of the DA method, which achieves high accuracy with considerably smaller reduced-order spaces. This favorable balance between accuracy and computational cost motivates our choice to focus on the DA method in the main text. Nevertheless, the PG method offers stronger theoretical guarantees on the accuracy of the spectral-strength moments. In particular, while the DA method guarantees exact moment matching at the snapshot parameter values, the PG method preserves the moments up to order $2\nu+1$ at arbitrary parameter values, within the accuracy of the EC approximation to the ground state. Consequently, although the DA method performs better for the Hubbard ring considered here, the PG approach may prove more advantageous in other systems or applications, particularly when accurate reproduction of higher-order spectral moments across the parameter space is essential.

\subsection{Time saved by employing the emulator}

In the main text, we reported the computational times obtained using Lanczos diagonalization. When full diagonalization is employed instead, the exact calculations require approximately 210 seconds, including 120 seconds for the ground-state calculations. By comparison, evaluating the ECGF takes only 1.7 seconds after a one-time emulator construction cost of 59 seconds, which is dominated by the large-scale ground-state diagonalizations.

Table~\ref{tab:TimeTaken} summarizes the computational times for both diagonalization approaches.

\begin{table}[h!]
    \centering
    \begin{tabular}{l|cc}
        \hline
        & Full diagonalization & Lanczos \\
        \hline
        Emulator construction & 59 & 6.7 \\
        ECGF evaluation & 1.7 & 1.3 \\
        Exact calculation & 210 & 8.4 \\
        \hline
    \end{tabular}
    \caption{Computational times (in seconds) required to construct and evaluate the emulator and to perform the exact calculations for the nine values of $U$ shown in Fig.~4 of the main text. Emulator construction is a one-time cost, while the evaluation times refer to all nine values used for $U$.}
    \label{tab:TimeTaken}
\end{table}

\clearpage

\bibliography{bibliography} 
\bibliographystyle{apsrev4-2}